\documentclass{elsart}
\usepackage{epsfig}

\def\gsim{\setbox0=\hbox{$>$}%
        \mathrel{\vtop{\baselineskip=0pt\lineskip=0pt%
        \copy0\hbox to\wd0{\hss$\scriptstyle\sim$}}}}

\begin{document}

\begin{frontmatter}
         
%
%%%%%%%%%%%%%%%%%%%%%%%%%%%%%%%%%%%%%%%%%%%%%%%%%%%%%%%%%%%%%%%%%%%%%%
\title{On the Seyfert 2 warm and cool infrared dichotomy}
%%%%%%%%%%%%%%%%%%%%%%%%%%%%%%%%%%%%%%%%%%%%%%%%%%%%%%%%%%%%%%%%%%%%%%
%
  
%%%%%%%%%%%%%%%%%%%%%%%%%%%%%%%%%%%%%%%%%%%%%%%%%%%%%%%%%%%%%%%%%%%%%%
\author{D.M. Alexander}
\address{SISSA, 2-4 via Beirut, 34014 Trieste, Italy}

%
%%%%%%%%%%%%%%%%%%%%%%%%%%%%%%%%%%%%%%%%%%%%%%%%%%%%%%%%%%%%%%%%%%%%%%
\begin{abstract}
%%%%%%%%%%%%%%%%%%%%%%%%%%%%%%%%%%%%%%%%%%%%%%%%%%%%%%%%%%%%%%%%%%%%%%
%

An optical spectropolarimetric study has shown that the detectability of
polarised broad H$\alpha$ in Seyfert 2 galaxies is correlated to the IRAS
$f_{60}/f_{25}$ flux ratio where only those Seyfert 2s with ``warm" IRAS
colours show polarised broad line emission. It was suggested that those
Seyfert 2s with ``cool" IRAS colours have highly inclined tori which
obscure the broad line scattering screen.

I present here hard X-ray observations inconsistent with this picture
showing that the derived column densities of warm and cool Seyfert 2
galaxies are statistically the same. I suggest that the IRAS
$f_{60}/f_{25}$ flux ratio is more consistent with implying the relative
strength of galactic to Seyfert emission and provide supporting evidence
for this view.

\end{abstract}


\end{frontmatter}

%
%%%%%%%%%%%%%%%%%%%%%%%%%%%%%%%%%%%%%%%%%%%%%%%%%%%%%%%%%%%%%%%%%%%%%%
\section{Introduction}
%%%%%%%%%%%%%%%%%%%%%%%%%%%%%%%%%%%%%%%%%%%%%%%%%%%%%%%%%%%%%%%%%%%%%%
%

The unified model for Seyfert galaxies proposes that all types of Seyfert
galaxy are fundamentally the same, however, the presence of a dusty
molecular ``torus" obscures the broad line region (BLR) in many systems.
In this picture the classification of Seyfert 1 or 2 (Seyfert 1--broad
permitted lines, Seyfert 2--narrow permitted lines) depends on the
inclination angle of the torus to the line of sight (Antonucci, 1993).
Probably the most convincing evidence for this unified model comes from
optical spectropolarimetry. Using this technique, the scattered emission
from the BLR of many Seyfert 2 galaxies is revealed in the form of broad
lines in the polarised flux (e.g.\ Antonucci and Miller, 1985, Young et
al, 1996, Heisler, Lumsden and Bailey, 1997).

In this unified picture the high energy central source emission (optical
to X-ray continuum) is absorbed by the dust within the torus which
re-emits this energy at infrared (IR) wavelengths. Independent strong
support has been given by hard X-ray (HX, 2 to 10 keV), near-IR and mid-IR
observations (e.g.\ Turner et al, 1997, Risaliti, Maiolino and Salvati,
1999, Alonso-Herrero, Ward, Kotilainen, 1997 and Clavel et al, 2000)  
showing that Seyfert 2 galaxies are generally characterised by strong
absorption whilst Seyfert 1 galaxies are relatively unabsorbed.

Heisler, Lumsden and Bailey (1997, hereafter HLB) performed an optical
spectropolarimetric study of a well defined and statistically complete
IRAS 60$\mu$m selected Seyfert 2 sample to determine the statistical
detectability of polarised broad lines. The objects were selected at
60$\mu$m to reduce the possibility of biasing due to torus
inclination/extinction effects and all objects were observed to the same
signal to noise to ensure similar detection thresholds. In this study a
striking relationship between the detectability of polarised broad
H$\alpha$ and the IRAS $f_{60}/f_{25}$ flux ratio was found where only
those galaxies with warm IRAS colours ($f_{60}/f_{25}<$4.0) showed a
hidden broad line region (HBLR). Both Seyfert 2 galaxy types were found to
be well matched in terms of redshift, overall polarisation and detection
rate of compact nuclear radio emission. Therefore, without any apparent
contradictory evidence, HLB suggested that the IRAS $f_{60}/f_{25}$ ratio
provides a measure of the inclination of the torus to the line of sight:
in a cool Seyfert 2 the torus is so highly inclined that even the broad
line scattering screen is obscured. I present here HX evidence that
strongly suggests that this picture is incorrect and provide a new view
that is consistent with other observations.

%
%%%%%%%%%%%%%%%%%%%%%%%%%%%%%%%%%%%%%%%%%%%%%%%%%%%%%%%%%%%%%%%%%%%%%%
\section{Testing the inclination picture}
%%%%%%%%%%%%%%%%%%%%%%%%%%%%%%%%%%%%%%%%%%%%%%%%%%%%%%%%%%%%%%%%%%%%%%
%

One of the key supports of the unified model come from HX observations
where the nuclear extinction is directly determined from the observed
spectral slope. Seyfert 1 galaxies are characterised by little or no
absorption 20$<$log($N_H$)$<$21 cm$^{-2}$ whilst Seyfert 2 galaxies have
significant, sometimes extreme, absorption 22$<$log($N_H$)$<$25 cm$^{-2}$
(e.g.\ Turner et al, 1997 and Risaliti, Maiolino and Salvati, 1999).  
According to the HLB interpretation the cool Seyfert 2s should show higher
column densities than the warm Seyfert 2s. To date 13 of the galaxies in
the HLB sample have been observed with either BeppoSAX or ASCA. The other
3 objects have been observed by Einstien or in the HEAO1/A survey. In the
case of the HEAO1/A objects only upper limits could be placed. For these
two galaxies (NGC34 and NGC1143) I have used the upper limits and
unextincted [OIII]$\lambda$5007 emission line fluxes to predict their
nuclear extinction using the diagnostic diagram of Bassani et al (1999).
The distribution of HX derived column densities are shown in figure 1.

%
%%%%%%%%%%%%%%%%%%%%%%%%%%%%%%%%%%%%%%%%%%%%%%%%%%%%%%%%%%%%%%%%%%%%%%
% Seyfert 2 NH distribution
%%%%%%%%%%%%%%%%%%%%%%%%%%%%%%%%%%%%%%%%%%%%%%%%%%%%%%%%%%%%%%%%%%%%%%
%

\begin{figure}
\begin{center}
\leavevmode
\centerline{\psfig{figure=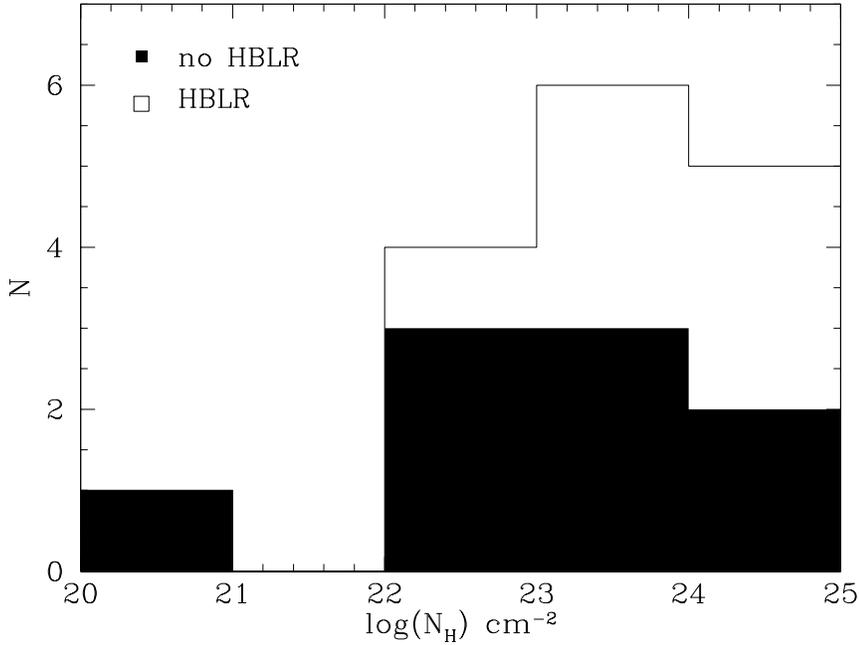,width=9cm,angle=-90}}
\vspace{0.5cm}
\end{center}
\caption{\em The distribution of hard X-ray derived hydrogen column
densities for the Seyfert 2 galaxies in the HLB sample.} 
\label{fig:sample1}
\end{figure}

%%%%%%%%%%%%%%%%%%%%%%%%%%%%%%%%%%%%%%%%%%%%%%%%%%%%%%%%%%%%%%%%%%%%%%

The derived column densities show that an optically thick structure exists
in both the warm and cool Seyfert 2 galaxy types although, signficantly,
there is very little difference in the distribution of column densities.
The only object which does not fit the general distribution is NGC7590
which may be a Seyfert 1 with galactic dust obscuring the BLR. The overall
distribution is similar to that found for the [OIII]$\lambda$5007 selected
Seyfert sample of Risaliti, Maiolino and Salvati (1999) suggesting that
the far-IR selects Seyferts in a reasonably unbiased manner: $\sim$35\% of
the objects are Compton thick (i.e. log($N_H)>24$ cm$^{-2}$), the mean
log($N_H$) for the whole sample is 23.2$\pm$0.9 cm$^{-2}$ and the mean for
the warm and cool Seyfert 2s are 23.7$\pm$0.5 cm$^{-2}$ and 22.9$\pm$1.0
cm$^{-2}$ respectively. It could be argued that the cool Seyfert 2s are
Compton thick and the determined column densities refer to the extinction
suffered by the scattered emission, however, the mean log([OIII]/HX) of
0.3$\pm$1.0 and -0.2$\pm$1.4 for the warm and cool Seyfert 2s respectively
suggest that this is not the case (see Bassani et al, 1999). If anything,
the cool Seyfert 2s appear to be HX bright compared to the warm Seyfert
2s.

The hypothesis of HLB could still be retained with some modification
(e.g.\ allowing different nuclear environments such as suggesting that the
warm Seyfert 2s have additional gaseous extinction within the torus walls
(Granato, Danese and Franceschini, 1997)). However, the simplest and most
direct conclusion is that the IRAS $f_{60}/f_{25}$ colour ratio does not
indicate the inclination angle of the torus in Seyfert 2 galaxies.

%
%%%%%%%%%%%%%%%%%%%%%%%%%%%%%%%%%%%%%%%%%%%%%%%%%%%%%%%%%%%%%%%%%%%%%%
\section{The Seyfert 2 infrared dichotomy}
%%%%%%%%%%%%%%%%%%%%%%%%%%%%%%%%%%%%%%%%%%%%%%%%%%%%%%%%%%%%%%%%%%%%%%
%

If the IRAS $f_{60}/f_{25}$ colour ratio is not an indicator of the
inclination of the dusty torus then what does this colour ratio imply? A
natural starting point is to compare the HLB Seyfert properties to those
of non-Seyfert galaxies. A good comparison is the Bright Galaxy Sample
(BGS, Soifer et al, 1989) which is selected at the same wavelength as the
HLB sample and has a very similar flux limit. The BGS sample is partially
classified by Kim et al (1995) using the optical emission line ratio
technique (e.g. Baldwin, Phillips and Terlevich, 1981). To increase the
number of classified objects I have taken these observations and other
optical spectroscopic observations from the literature, classifying 77\%
of the BGS sample: 25\% are found to be LINERs, 62\% are HII galaxies,
12\% are AGN and 1\% have no emission lines. These galaxies have been
classified using all the emission line diagnostics of Veilleux and
Osterbrock (1987) and the mode classification for each galaxy is adopted.
For brevity only the [NII]$\lambda$6583/H$\alpha$ vs
[OIII]$\lambda$5007/H$\beta$ diagram is shown here, see figure 2.
 
%
%%%%%%%%%%%%%%%%%%%%%%%%%%%%%%%%%%%%%%%%%%%%%%%%%%%%%%%%%%%%%%%%%%%%%%
% Emission line distribution
%%%%%%%%%%%%%%%%%%%%%%%%%%%%%%%%%%%%%%%%%%%%%%%%%%%%%%%%%%%%%%%%%%%%%%
%

\begin{figure}
\begin{center}
\leavevmode
\centerline{\psfig{figure=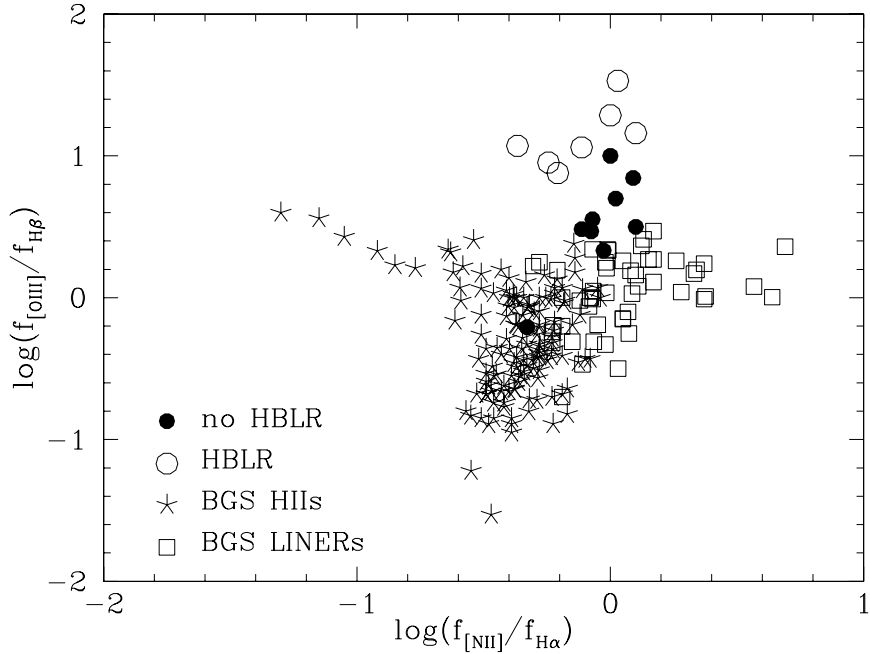,width=9cm,angle=-90}}
\vspace{0.5cm}
\end{center}
\caption{\em The optical emission line ratios for LINERs and HII galaxies
from BGS and Seyfert 2 galaxies from HLB.}
\label{fig:sample1}
\end{figure}

%%%%%%%%%%%%%%%%%%%%%%%%%%%%%%%%%%%%%%%%%%%%%%%%%%%%%%%%%%%%%%%%%%%%%%

A logical first step is to compare the IRAS colours, see figure 3. The IR
warm region is clearly dominated by Seyfert 2s although the cool region
also contains HII and LINER galaxies with a wide range of IRAS colours.
The cool Seyfert 2s cannot be distinguished from the HII and LINER
galaxies in terms of their IRAS emission (note there are no Seyfert 2s
with log($f_{60}/f_{25})>$0.9 due to the HLB selection criteria). As the
distribution of HX derived nuclear column densities are so similar it is
probable that the same optically thick structure (i.e.\ the torus) exists
in both the warm and cool Seyfert 2 galaxy types. According to the unified
model, this structure should emit thermally at IR wavelengths and
therefore, although the HX emission shows that a Seyfert nucleus is
present in the cool Seyfert 2s, the IR emission from the torus must be
dominated by galactic emission in the large IRAS apertures (as previously
suggested by Alexander et al, 1999). Additional evidence for this picture
is found in the distribution of optical emission line ratios where the
cool Seyfert 2s have, on average, weaker [OIII]/H$\beta$ emission, see
figure 2. Assuming that both the warm and cool Seyfert 2s have the same
basic Seyfert nucleus and galactic emission, the lower mean emission line
ratio in the cool Seyfert 2s implies a larger ratio of galactic to Seyfert
activity. Indeed in one galaxy (NGC7496) the observed emission line ratio
is consistent with that of an HII galaxy even though it clearly has HX
emission and therefore a Seyfert nucleus.

%
%%%%%%%%%%%%%%%%%%%%%%%%%%%%%%%%%%%%%%%%%%%%%%%%%%%%%%%%%%%%%%%%%%%%%%
% IR distribution of galaxies
%%%%%%%%%%%%%%%%%%%%%%%%%%%%%%%%%%%%%%%%%%%%%%%%%%%%%%%%%%%%%%%%%%%%%%
%

\begin{figure}
\begin{center}
\leavevmode
\centerline{\psfig{figure=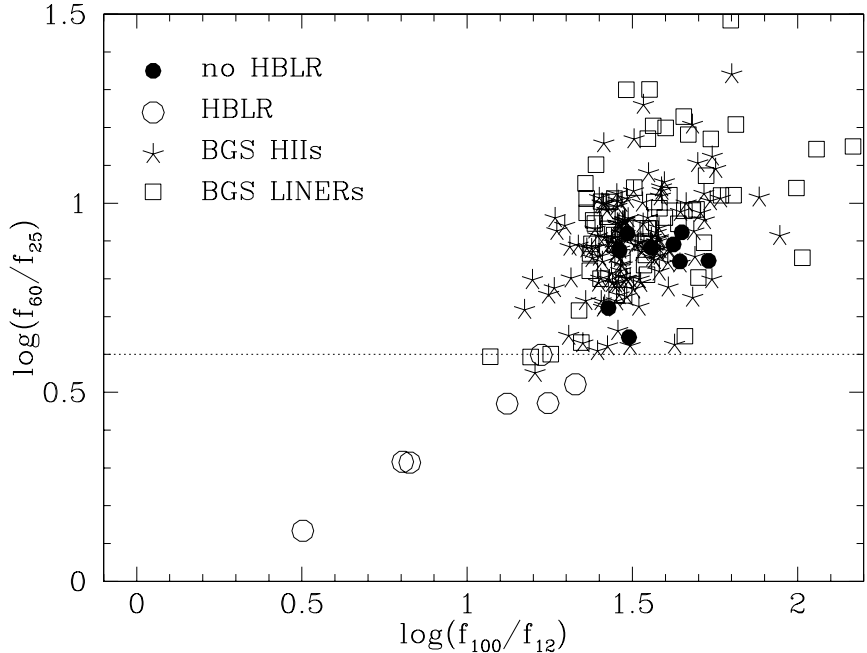,width=9cm,angle=-90}}
\vspace{0.5cm}
\end{center}
\caption{\em The IRAS colour distribution for LINERs and HII galaxies from
BGS and Seyfert 2 galaxies from HLB. The warm/cool divide is indicated by
the dotted line.}
\label{fig:sample1}
\end{figure}

%%%%%%%%%%%%%%%%%%%%%%%%%%%%%%%%%%%%%%%%%%%%%%%%%%%%%%%%%%%%%%%%%%%%%%

%
%%%%%%%%%%%%%%%%%%%%%%%%%%%%%%%%%%%%%%%%%%%%%%%%%%%%%%%%%%%%%%%%%%%%%%
\section{Conclusions}
%%%%%%%%%%%%%%%%%%%%%%%%%%%%%%%%%%%%%%%%%%%%%%%%%%%%%%%%%%%%%%%%%%%%%%
%

I have presented HX observations of Seyfert 2 galaxies that are
inconsistent with the HLB explanation for the IR dichotomy. From the
classification of the BGS sample I have shown that the distribution of
IRAS colours and optical emission line ratios favour the IRAS
$f_{60}/f_{25}$ flux ratio implying the strength of galactic to Seyfert
activity. I provide further evidence for this view and explain the
spectropolarimetric results in a more detailed article (Alexander, 2000).

\section*{Acknowledgements} I am in debt to the late Dr. Charlene
Heisler for interesting discussions whilst working on this topic. Her
enthusiasm and contribution to astronomy will be greatly missed.

%
%%%%%%%%%%%%%%%%%%%%%%%%%%%%%%%%%%%%%%%%%%%%%%%%%%%%%%%%%%%%%%%%%%%%%%

%%%%%%%%%%%%%%%%%%%%%%%%%%%%%%%%%%%%%%%%%%%%%%%%%%%%%%%%%%%%%%%%%%%%%%


\begin{thebibliography}{99}
%%%%%%%%%%%%%%%%%%%%%%%%%%%%%%%%%%%%%%%%%%%%%%%%%%%%%%%%%%%%%%%%%%%%%%
%

\bibitem{}
Alexander, D.M., Efstathiou, A., Hough, J.H., Aitken, D.K., Lutz, D.,
Roche, P.F., Sturm, E.\ 1999, MNRAS, 310, 78

\bibitem{}
Alexander, D.M.\ 2000, MNRAS, submitted

\bibitem{}
Alonso-Herrero, A., Ward, M.J., Kotilainen, K.\ 1997, MNRAS, 288, 977

\bibitem{}
Antonucci, R., Miller, J.\ 1985, ApJ, 297, 621

\bibitem{}
Antonucci, R.\ 1993, ARA\&A, 31, 473
 
\bibitem{}
Baldwin, J.A., Phillips, M.M., Terlevich, R. \ 1981, PASP, 93, 5

\bibitem{}
Bassani, L., et al.\ 1999, ApJS, 121, 473

\bibitem{}
Clavel, J., et al.\ 2000, A\&A, in press

\bibitem{}
Granato, G.L., Danese, L., Franceschini, A.\ 1997, ApJ, 486, 147

\bibitem{}
Heisler, C.A., Lumsden, S.L., Bailey, J.A.\ 1997, Nature, 385, 700 (HLB)

\bibitem{}
Kim, D.-C., Sanders, D. B., Veilleux, S., Mazzarella, J. M., Soifer, B.
T.\ 1995, ApJS, 98, 129

\bibitem{}
Risaliti, G., Maiolino, R., Salvati, M.\ 1999, ApJ, 522, 157

\bibitem{}
Soifer, B.T., Boehmer, L., Neugebauer, G., Sanders, D.B.\ 1989, AJ, 98,
766 (BGS)

\bibitem{}
Turner, T.J, George, I.M., Nandra, K. \& Mushotzky, R.F. 1997, ApJ, 488, 164
 
\bibitem{}
Veilleux, S., Osterbrock, D.\ 1987, ApJS, 63, 295

\bibitem{}
Young, S., Hough, J.H., Efstathiou, A., Wills, B.J., Bailey, J.A., Ward,
M.J., Axon, D.J.\ 1996, MNRAS, 281, 1206


\end{thebibliography}
\end{document}